# A flexible and scalable single-level framework for OD matrix inference using multiple sources of transport information

27 September 2022

Wei Sun (corresponding author)
Research Fellow
University of South Australia
City West Campus (WL5-39) Adelaide, SA 5000
 weisun.simon@gmail.com
+61 4 5252 0309

Akshay Vij (corresponding author)
Associate Professor
University of South Australia
City West Campus (WL5-41) Adelaide, SA 5000
vij.akshay@gmail.com
+61 8 8302 0817

Nicolas Kaliszewski
Director, Transport Analytics
South Australian Department for Infrastructure & Transport
77 Grenfell Street Adelaide, SA 5000
Nicholas.Kaliszewski@sa.gov.au
+61 4 6684 2754

**Abstract**

This study proposes a flexible and scalable single-level framework for origin-destination matrix (ODM) inference using data from IoT (Internet of Things) and other sources. The framework allows the analyst to integrate information from multiple data sources, while controlling for differences in data quality across sources. We assess the effectiveness of the framework through a real-world experiment in Greater Adelaide (GA), Australia. We infer car OD flows within the region using four separate data sources: site-level traffic counts from loop detectors, vehicle trajectories recorded by roadside Bluetooth sensors, partial OD flows based on data from in-vehicle navigation systems, and journey-to-work data collected by the Australian Census. We compare our OD inferences with those from the current version of the Metropolitan Adelaide Strategic Transport Model (MASTEM), calibrated using data from traditional household travel surveys. We find remarkable consistency between our inferences and those from MASTEM, despite differences in input data and methodologies. For example, for the morning peak period, we predict the total number of trips made within GA to be equal to 556,000, while the corresponding prediction from MASTEM is equal to 484,000. The two predictions are within 20 per cent of each other. When we compare the spatial distribution of trips, in terms of origins and destinations, we find that our inferred OD matrix has an 86 per cent cosine similarity to the corresponding MASTEM matrix. In summary, our results show that the proposed framework can produce highly comparable ODM, trip production and trip attraction patterns to those inferred from traditional household travel survey-based transportation demand modelling methods.

# 1. Introduction

In transport engineering and planning, the origin-destination matrix (ODM) refers to the distribution of trips in terms of their origins and destinations between zones within a planning area. The ODM provides insights on travel patterns within the planning area, and is therefore one of the key outputs of the strategic transport modelling and forecasting process. Traditionally, ODMs have been inferred using data collected through household travel surveys (HTS) that ask participating household members about their travel patterns over a 1 or 2- day observation period. However, HTSs are expensive. Hartgen and San Jose (2009) report average costs of $487,000 per HTS, and roughly $150 per response, though they note that "many surveys cost considerably more than the average, and the spread of the data is substantial". Stopher et al., (2011) find that a computer-assisted telephone interview survey in Australia would cost $150-200 per household, face-to-face surveys are likely in the order of $350 plus per household, and a 15-day GPS survey would cost around $300 per household. Furthermore, HTSs are typically cross-sectional, only providing a static snapshot, and most urban areas around the world conduct their HTS only once every 5-10 years. Older surveys can lose their representativeness, especially if there are structural shifts in travel patterns. For example, the recent COVID-19 pandemic and accompanying public health concerns are likely to have long-term impacts on trip-making and mode-use patterns that would not be captured by surveys conducted before the pandemic. Similarly, the emergence of new modes of transport, such as electrical scooters recently, and ridesharing in the recent past, have had similar if smaller impacts on travel patterns that may not be fully captured by older surveys.

The rapid diffusion of smartphones, Wi-Fi and Bluetooth networks, and the digitization of transport planning, booking and payment systems, in conjunction with broader advances in the Internet of Things (IoT) across all sectors of the economy, imply that we have more data than ever before on how people use transport infrastructure, and how these patterns are likely to change in the future. These new ICT technologies offer a more cost-effective alternative for the collection of transport data, but at far greater volumes. For example, the Sydney Household Travel Survey currently samples roughly 5,000 households each year from a population of roughly 5 million, and observes their travel patterns over a 24-hour period. In contrast, data from IoT sources could offer a continuous stream of travel information for a large majority of that population over rolling time periods.

As a result, in recent decades, numerous studies have developed methods for the inference of ODMs based on data collected from vehicle loop detectors, Bluetooth detectors, license plate scanners, GPS-connected devices and other IoT sources. Spiess (1990) proposed an approach to estimate the ODM using observed link flows, and his study has paved the way for subsequent studies in this area. Apart from traffic counts data, with increased data availability, subsequent studies have utilized other IoT sources, such as mobile phone data, vehicle turning data and Bluetooth data, to estimate the ODM. These data sources have either been used independently to infer the ODM (Carpenter et al., 2012, Barceló et al., 2013, Alexander et al., 2015), or used jointly with traffic counts from loop detectors to improve inference accuracy (Iqbal et al., 2014, Behara et al., 2021a, Behara et al., 2021b, Michau et al., 2019, Cipriani et al., 2021).

A review of the literature suggests there are two key limitations to these previous studies that the present study aims to overcome. First, existing studies estimate the ODM based on data from only one or two IoT sources. It is not clear how information from new sources can be integrated within their frameworks, and if and how the analyst can control for differences in quality across datasets. We develop a flexible framework that allows the analyst to integrate multiple data sources for ODM inference, and to ascribe weights to each dataset as a measure of their quality to be used accordingly within the ODM inference algorithm. For example, in our real-world experiment, we use information from vehicle loop detectors, Bluetooth sensors, in-vehicle navigation systems and commute mode choice data collected by the national Census to infer ODMs in Adelaide, Australia. During inference, we give the greatest weight to roadside loop detector data and the lowest weight to data from in-vehicle navigation systems, to capture known differences in observation quality.

Second, many studies have followed a bi-level framework that uses an equilibrium or optimal model to perform path assignment, i.e., to translate ODMs into equivalent link counts. More recently, studies such as Dey et al. (2020) and Behara et al. (2021b) propose to replace the traditionally equilibrium-based path choice assignment with inferred or pre-determined path choice functions based on historic data. By using the inferred path choice function from the observed data, Behara et al. (2021b) simplify the traditional bi-level framework to a single-level, which is much more computationally efficient, especially when applying to a large road network. Therefore, instead of testing ODM inference quality on a relatively small part of the city/region with only a few hundred OD pairs (Cipriani et al., 2021, Behara et al., 2021a, Ma and Qian, 2018, Cantelmo et al., 2014), the framework can be easily deployed to a city-wide scale (Behara et al., 2021b). For example, our real-world experiment covers the Adelaide metropolitan region in Australia, comprising roughly 1.3 million residents living and working across 3,260 square-km, and contains ten thousand OD pairs. By adopting a single-level framework, we can scale our algorithms to infer ODMs for the entire region.

In summary, we propose a flexible and scalable algorithm for OD inference, which can leverage travel information from multiple sources with varying degrees of quality, and which can be applied to large planning areas at a network-wide and area-wide level. The framework offers a viable alternative to traditional survey-based methods of ODM inference.

The reminder of this paper is structured as follows. Section 2 provides some background and context to our case study region of Adelaide in order to motivate the general ODM inference problem. Section 3 develops a framework for ODM inference using passively collected movement and count data from different sources. Section 4 applies this framework to infer car OD flows in Adelaide using data from vehicle loop detectors, Bluetooth sensors, in-vehicle navigation systems and the Census. Section 5 undertakes a robustness analysis to examine the sensitivity of our inference algorithms to different input parameters. Section 6 concludes with a summary of the key findings and directions for future research.

## 2. Background and context

The city of Adelaide is the capital of the state of South Australia, and the fifth largest Australian city. It has a population of roughly 1.3 million and an area of 3260 km$^2$. The South Australian Department for Infrastructure & Transport (DIT) is responsible for the delivery of effective planning policy, efficient transport, and valuable social and economic infrastructure in Adelaide. The performance of strategic transport models currently being used by DIT has been undermined by limited resources. In particular, the current Metropolitan Adelaide Strategic Transport Evaluation Model (MASTEM) - DIT's strategic travel demand model for Adelaide – was calibrated using data from the 1999 Metropolitan Adelaide Household Travel Survey. That data is now more than twenty years old, and not reflective of current or future travel patterns within the region. There is an urgent need to update the model using current data. This study is part of a broader project investigating the ability of passively-collected transport data from different IoT and other sources to replace traditional household travel survey data collection methods for these purposes.

The objective of this study is to develop a general methodological framework for inferring OD flows using transport data from multiple existing information sources with varying levels of data quality. In the first instance, we apply this framework to infer car OD flows within Adelaide, at a spatial resolution corresponding to Statistical Area Level 2 (SA2) zones, segmented by time-of-day. Adelaide comprises 102 SA2 zones, and these are defined as the origin and destination zones for our OD inference problem. The average SA2 in Adelaide has a population of 12000 residents and an area of 18.35 km$^2$, and is roughly ten times larger than the average traffic analysis zone (TAZ). Ideally, we wish to infer OD flows at the TAZ-level. However, at present, our information sources are not dense enough to allow that level of spatial analysis. As part of ongoing work, we are exploring ways in which findings from the present study could be interpolated to produce analogous ODMs at the TAZ-level. Similarly, we are examining ways in which our framework can be applied to infer OD flows for other modes, such as public transport and bicycling. However, a detailed discussion of either topic is considered beyond the scope of the present study.

We describe our data sources in more detail over following paragraphs. First, we use information collected by 1003 road-side Bluetooth sensors installed across the Adelaide road network. These sensors are able to detect active Bluetooth devices within a 90 m radius of the sensor, and able to track device movements across the network using their unique MAC address. However, as shown in Figure 1, many parts of the network, especially in the peripheral regions of the metropolitan area, are not covered by sensors. Additionally, only vehicles with an active Bluetooth device are detected by these sensors, and these are estimated to comprise roughly 10-15 per cent of all road traffic.

Second, we use information collected by 835 loop detectors installed at signalized intersections across the Adelaide road network, as part of the network's Sydney Coordinated Adaptive Traffic System (SCATS). These loop detectors are able to record the number of total vehicles passing through each lane at any signalised intersection within the network over any time period, excluding left-turning movements. However, they do not track individual vehicle movements through the network. Furthermore, as shown again in Figure 1, not all road intersections are signalised and have loop detectors.

Third, we use information collected by in-vehicle navigation systems that use map data provided by HERE Technologies. This information is available in the form of aggregated time-stamped OD flows at an SA2 level. However, only about 10 per cent of the total vehicle fleet in Adelaide is estimated to use HERE Technologies' mapping data and navigation systems, and similar information is not available for other vehicle makes and models.

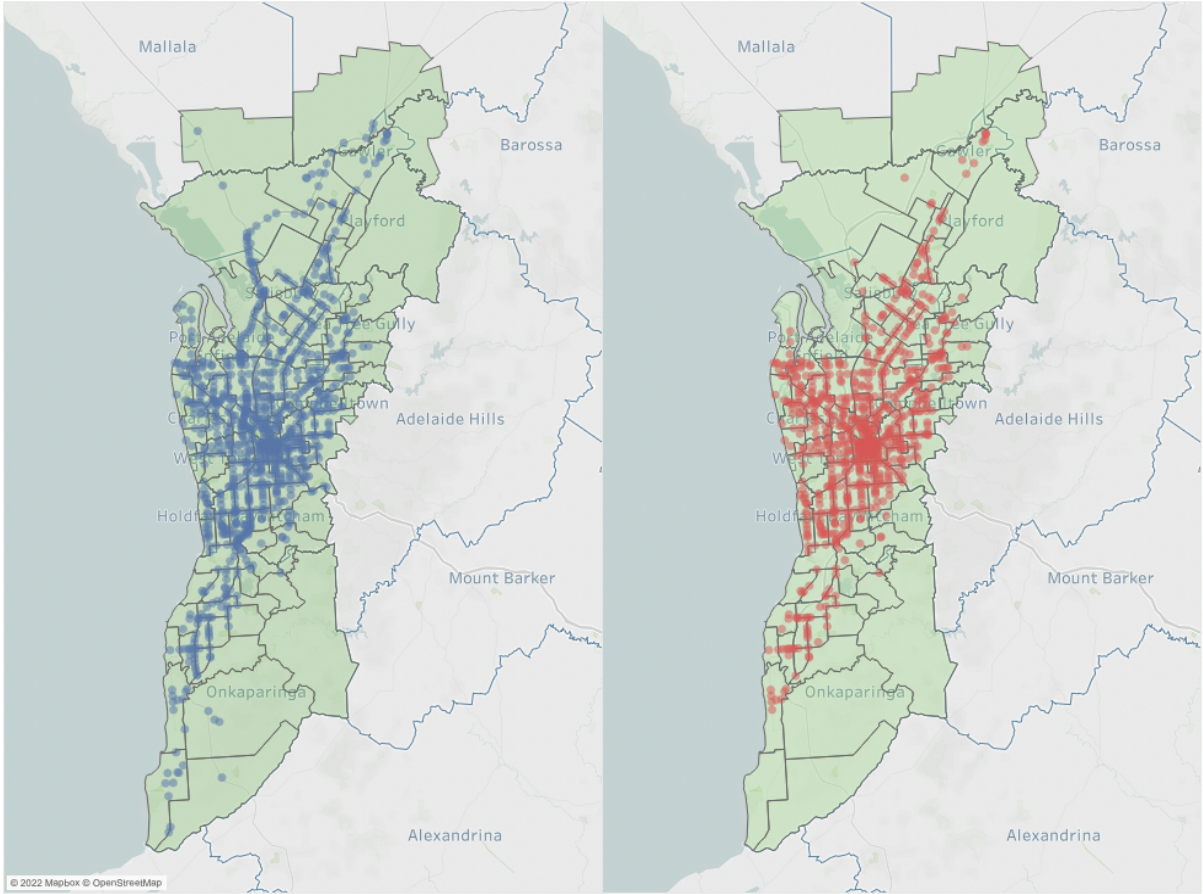

**Figure 1:** Location of Bluetooth sensors (left) and loop detectors (right) across the Adelaide metropolitan area; the black lines denote SA2 boundaries

Fourth, we use information collected by the Australian Census. The Census is conducted once every five years. All working individuals are asked to report the location of their place of residence and place of work, and the transport mode used to commute to work on Census day. This information can be used to generate car OD flows for work travel on Census day. However, this information is not available at a greater temporal resolution (e.g. hourly), and this information is also not available for non-work travel.

In general, each of these data sources offers a partial view of car OD flows across the network. In the following section, we develop a general optimisation framework that infers the OD matrix that is most likely to produce these partial views observed across these different data sources.

## 3. Methodological framework

We develop a general methodological framework for inferring OD flows using transport data from multiple sources with varying levels of data quality. While the framework has been developed to infer car OD flows using the data sources described in the previous section, it is general in that it can easily incorporate any movement or count information from other sources, such as license plate scanners, GPS devices, smartphones, smartcards, surveys, traffic counters, etc.

Transport information collected by these different data sources can broadly be categorised into two types. First, sources such as Bluetooth sensors, in-vehicle navigation devices and Census surveys provide partial ODM information either directly, or indirectly in the form of observed network movements or trajectories that can be converted into analogous OD flows at a zone-level. Note that these sources only offer a partial view of OD flows within the network, and we need to combine information from multiple sources to construct the full ODM. For example, Bluetooth sensors only cover a fraction of the road network in any metropolitan area, and only vehicles with an active Bluetooth device are detected by these sensors. In-vehicle navigation systems provide a higher level of spatial resolution for observed trips, but information from these systems is only available for a subset of vehicle makes and models. The Australian Census provides an estimate of OD flows for work trips on Census day by different modes of transport, but does not offer greater temporal resolution, and this information is not available for non-work travel. Depending on the quality of the data source, the analyst may have to develop supplementary algorithms to convert the raw information into comparable OD flows. We denote observed partial OD flows based on different data sources by the matrix $\mathbf{T_j}$, where $j$ denotes the data source, and $\mathbf{T_j}$ is an $(R_j \times R_j)$ matrix whose element $t_{jmn}$ denotes observed flows between origin zone $m$ and destination zone $n$ based on data source $j$, and $R_j$ denotes the number of zones reliably covered by data source $j$. Note that $R_j$ might be smaller than the total number of zones in the planning region. For example, as mentioned previously, there are 1003 Bluetooth sensors across the Adelaide road network, and together they cover 94 of the 102 SA2 zones within the Adelaide metropolitan region. Further, let $F_j$ denote the function used to derive OD flows across the partial network covered by data source $j$, from the OD flows for the full network in the planning area.

Second, sources such as vehicle loop detectors and traffic counters provide count data, such as the number of vehicles that pass through a particular intersection, or the number of passengers that alight from a public transport vehicle at a particular stop. These sources do not offer any information on network movements or trajectories. For example, with loop detectors at signalised intersections, we do not know where the vehicles are coming from or where they are going to. In order to use count information for ODM inference, we need to develop supplementary algorithms to translate a given ODM into comparable counts at different counter sites. Let $G_k$ denote the function used to derive counts across the sites covered by data source $k$, from the OD flows for the full network in the planning area. Further, let $\mathbf{q_k}$ denote different sources of count data, where $k$ denotes the count data source, and $\mathbf{q_k}$ is an $S_k$-dimensional vector whose element $q_{kv}$ denotes observed counts at site $v$, and $S_k$ is the number of independent sites corresponding to the data source. For example, for our Adelaide case study, we use information from 781 loop detector sites.

Given a set of different data sources that offer a partial view of OD flows and/or counts across the network, we want to develop a flexible optimisation framework that infers the OD matrix that is most likely to produce these partial views across these different data sources, while controlling for differences in data quality across these different sources. Mathematically, we formulate this optimisation problem as follows:

$$\widehat{\mathbf{B}} = \underset{\mathbf{B}}{\text{argmin}} \left[ \sum_j w_j \cdot c_j(F_j(\mathbf{B}), \mathbf{T_j}) + \sum_k w_k \cdot d_k(G_k(\mathbf{B}), \mathbf{q_k}) \right] \quad (1)$$

, where $\mathbf{B}$ is the OD matrix for the full network in the planning area, and $\widehat{\mathbf{B}}$ is our estimate for the same; $w_j$ and $w_k$ are weights between 0 and 1 assigned a priori by the analyst denoting the quality of movement data source j and count data source k, respectively; $F_j(\mathbf{B})$ denotes the inferred partial ODM over the subset of the planning area covered reliably by data source j, and $G_k(\mathbf{B})$ denotes the inferred traffic counts at counter sites corresponding to data source k; and $c_j$ and $d_k$ denote the corresponding distance functions for these data sources that measure the degree of divergence between our inference and observed data.

The optimization algorithm minimises the weighted sum of the divergence between the inferred OD matrix on one hand, and observed movement and count data on the other. Given that the divergence is summed across different data sources, it is important that the distance functions are chosen to be on the same scale, so that no one data source dominates the optimisation function.

For movement data sources, we measure the distance between observed OD flows $\mathbf{T_j}$ and inferred OD flows $F_j(\mathbf{B})$ in terms of the cosine distance, where we flatten the matrices and treat them as vectors, defined as follows:

$$c_j(F_j(\mathbf{B}), \mathbf{T_j}) = 1 - \frac{\sum_m \sum_n \hat{t}_{jmn} \cdot t_{jmn}}{\sqrt{\sum_m \sum_n \hat{t}_{jmn}^2} \cdot \sqrt{\sum_m \sum_n t_{jmn}^2}} \quad (2)$$

, where $\hat{t}_{jmn}$ is the $(m, n)$ element of the matrix $F_j(\mathbf{B})$. Note that most data sources only record a fraction of total travel in the planning area. For example, for our case study context of Adelaide, Bluetooth sensors are estimated to record roughly 10-15 per cent of all passing traffic, data for in-vehicle navigation systems only covers OD flows for 10 per cent of the total vehicle fleet, and the Census only offers insights on OD flows for work travel (which typically comprise 30 per cent of total travel in large cities). Therefore, OD flows observed by these different sources in terms of absolute trip numbers are likely to vary substantially in scale from total inferred OD flows. However, the spatial distribution of observed trips in relative terms across these data sources is likely still to offer valuable information on the spatial distribution of all trips within the planning area. Therefore, we use the cosine distance measure that both allows us to include this structural information within our framework, while controlling for differences in scale across different data sources.

For count data sources, we calculate the Euclidean distance between the vectors $G_k(\mathbf{B})$ and $\mathbf{q_k}$, and normalise it between 0 and 1 to make it comparable in scale to the cosine distance for movement data sources:

$$d_k(G_k(\mathbf{B}), \mathbf{q_k}) = \frac{\sqrt{\sum_v (q_{kv} - \hat{q}_{kv})^2}}{\sqrt{\sum_v (\hat{q}_{kv})^2}} \quad (3)$$

, where $\hat{q}_{kv}$ is the $v^{th}$ element of the vector $G_k(\mathbf{B})$.

We still need to specify the functions $F_j$ and $G_k$. The former is relatively straightforward in that it extracts the subset of the full ODM that corresponds to the zones covered by data source j. For the latter, two broad approaches have been employed in the literature. Some studies have followed a bi-level framework that uses an equilibrium model to perform path assignment (Cipriani et al., 2021, Behara et al., 2021a, Ma and Qian, 2018, Cantelmo et al., 2014). Trips are assigned to routes based on a decision function (e.g. shortest path, utility maximisation), link-level travel times are updated based on the distribution of assigned trips, the new travel times are fed back to the path assignment model to update link-level flows, and the process is repeated until link-level travel times and flows converge to a stable equilibrium. We propose using the single-level framework developed by studies such as Dey et al. (2020) and Behara et al. (2021b) that replaces the equilibrium-based path choice assignment with inferred or pre-determined path choice functions based on historic data. A static function is developed that assigns trips for each OD pair to different routes in proportion to their historic use. In our case, this information can readily be derived from the Bluetooth data.

In words, our optimisation problem searches for the full OD matrix $\mathbf{B}$ that is most likely to produce the partial movement data $\mathbf{T_j}$ and count data $\mathbf{q_k}$ observed across different data sources $j$ and $k$. The optimisation problem is non-convex, and likely to have multiple local minima. Consequently, estimation results could potentially be sensitive to starting values, such that the inferred ODM does not reach the global optimum, and the optimisation algorithm ends up at a local optimum. Therefore, we recommend repeating inference with different starting values, and assessing the impact on estimation results. We conduct a similar sensitivity test as part of our case study, and the process is described in greater detail in Section 5. All inferences reported in this paper were implemented in Python using the Pyomo[1] library and the ipopt solver[2].

The proposed framework is flexible in that it can incorporate any kind of movement or count data for any transport mode. For example, the framework could easily be adapted to infer OD flows for bicycling, using movement data from surveys and bikeshare systems, and count data from bicycle counters. Unlike frameworks proposed by previous studies that can only use one or two sources of information (Carpenter et al., 2012, Alexander et al., 2015, Iqbal et al., 2014, Michau et al., 2019), our framework can readily incorporate multiple sources. Finally, the weights $w_j$ and $w_k$ in (1) allow the analyst to control for differences in data quality. For example, in our application, we assign a higher weight to data from loop detectors than Bluetooth sensors, as loop detectors are likely to capture a higher proportion of passing vehicles than Bluetooth sensors, and therefore likely to be a more reliable source of information.

The proposed framework is scalable in that it can readily be operationalised for large realistic networks and urban areas. For example, as we demonstrate in the next section, we were able to apply the framework to infer OD flows across an urban area comprising 104 SA2 zones, or roughly ten thousand distinct OD pairs. This is because we adopted a single-level framework for path assignment. Unlike the bi-level framework, the process is not iterative, and therefore much more computationally efficient. In their comparison of the two approaches, Behara et al. (2021b) find that the single-level framework is not only computationally faster, but it "yielded slightly better results than the traditional bi-level [framework]." For more details on the two frameworks and how they compare, the reader is referred to Behara et al. (2021b).

---

[1] About Pyomo. http://www.pyomo.org/about
[2] Ipopt documentation. https://coin-or.github.io/Ipopt

## 4. Inference results for case study

We apply the framework presented in the previous section to infer OD flows for car travel in Adelaide, Australia. As mentioned previously, the metropolitan area comprises 102 SA2 zones, and these are defined as the origin and destination zones for our OD inference problem. We use partial ODM data from Bluetooth sensors, in-vehicle navigation systems and the Australian Census, and vehicle count data from the SCATS system, for inference. We focus our attention on inferring flows during the AM peak period (7 – 10 AM) for a typical weekday in May 2019. We have applied the algorithm to other time periods, and the results are largely similar. For the sake of brevity, we do not include them here.

Figure 2 presents the pre-processing workflow to convert raw data into inputs into the inference algorithm. Partial ODM data from both the Census and HERE Technologies' in-vehicle navigation systems was provided directly. Data from the Bluetooth sensors and SCATS sites required significant pre-processing. Raw data from Bluetooth sensors was converted to vehicle trajectory data. This data was spatially aggregated to derive partial ODMs from the Bluetooth sensors. This data was also used to develop a path choice model to help convert ODMs into comparable SCATS site-level counts. We also did some quality checks by comparing data from individual Bluetooth sensors and SCATS sites in close proximity to each other. If we detected a significant difference in observed flows between the two, we excluded these Bluetooth sensors and SCATS sites from our analysis.

The processed data is subsequently input into our optimisation framework to infer the ODM for the full planning area. Note that our framework requires the analyst to assign weights between 0 and 1 to each data input, based on the quality of information. In our case, we assigned a weight of 1.00 to the SCATS data, 0.75 to the Census data, 0.50 to the Bluetooth data, and 0.30 to the HERE Technologies data. SCATS counters capture a large proportion of traffic flows (all vehicles passing through a signalised intersection except for left-turning movements), and are therefore given the highest weight. Conversely, HERE Technologies are estimated to capture less than 10 per cent of total vehicle movements in the area, and are therefore given the lowest weight. In the following section, we undertake a sensitivity analysis to assess the impact of different weights on our inference results.

Based on these weights, we estimate the ODM, and we compare it to the ODM for the same time period as estimated by MASTEM based on inputs from a traditional household travel survey. In summary, we find remarkable consistency between our inferences and those from MASTEM, despite differences in input data and methodologies. For example, we predict the total number of trips made within GA to be equal to 556,000, while the corresponding prediction from MASTEM is equal to 484,000. The two predictions are within 20 per cent of each other. When we compare the spatial distribution of trips, in terms of origins and destinations, we find that our inferred OD matrix has an 86 per cent cosine similarity to the corresponding MASTEM matrix.

Figure 3 compares trip production rates for the AM peak period across different zones. Visually, there is a high degree of similarity between the maps. The cosine similarity between our inferred ODM and the ODM from MASTEM is 0.97, and the correlation coefficient is 0.85. Analogously, Figure 4 plots the number of trips produced by the top twenty zones in the GA region. Again, for a majority of the zones, predictions from our inference algorithm are within 10 per cent of the corresponding predictions from MASTEM. For trip attractions, the results are even more closely aligned. The cosine similarity between the two OD matrices is 0.98, and the correlation coefficient 0.97. For the sake of brevity, we do not include analogous charts for trip attraction.

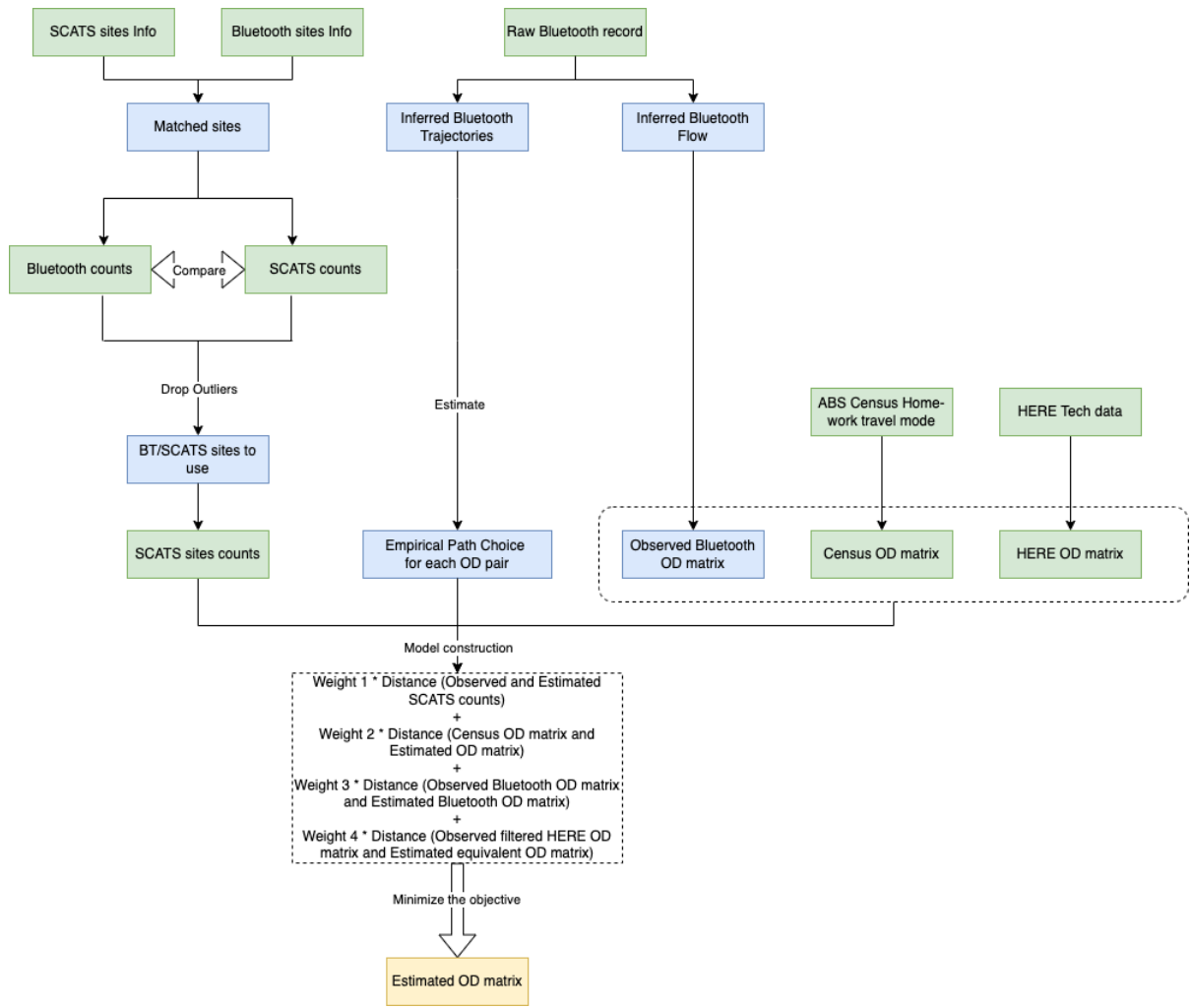

**Figure 2:** Data sources and data pre-processing workflow

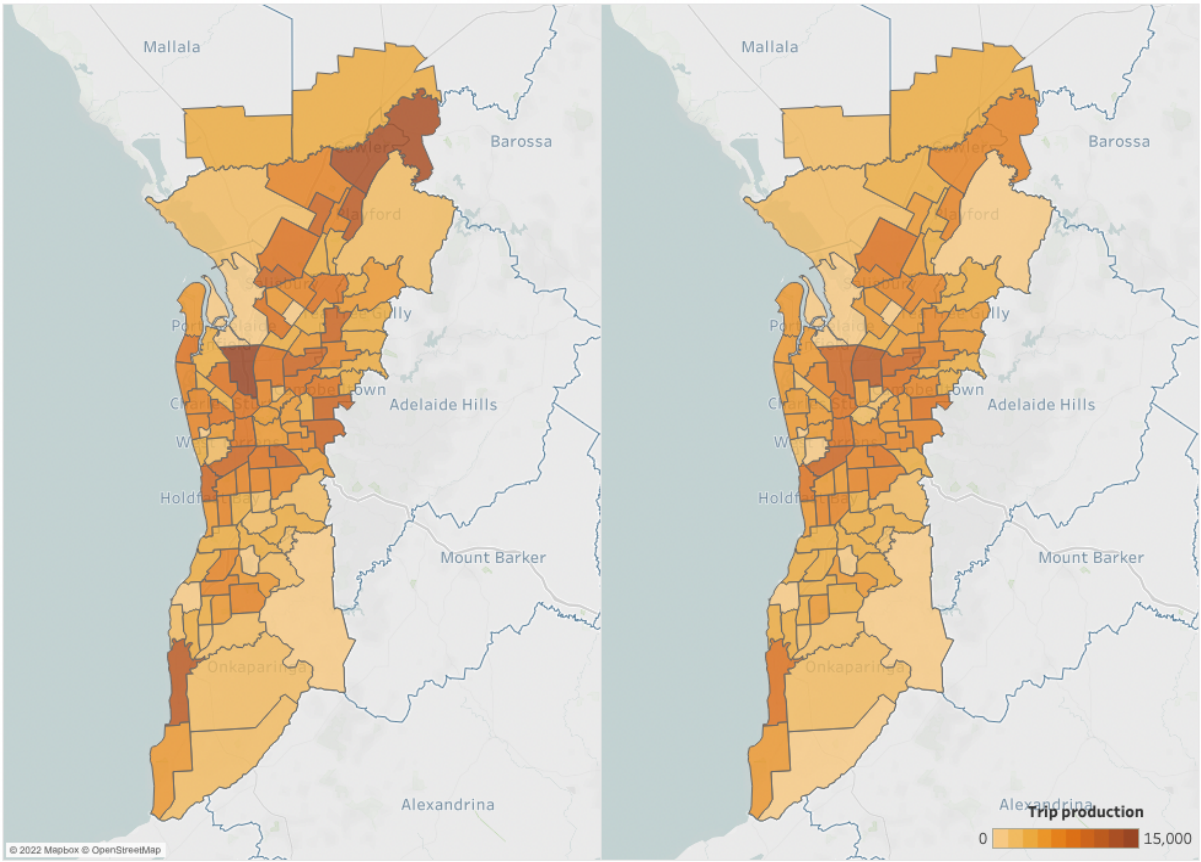

**Figure 3:** Comparison between zone-level trip production rates for the weekday AM peak period, as predicted by our inference algorithm (left), and the current version of MASTEM (right)

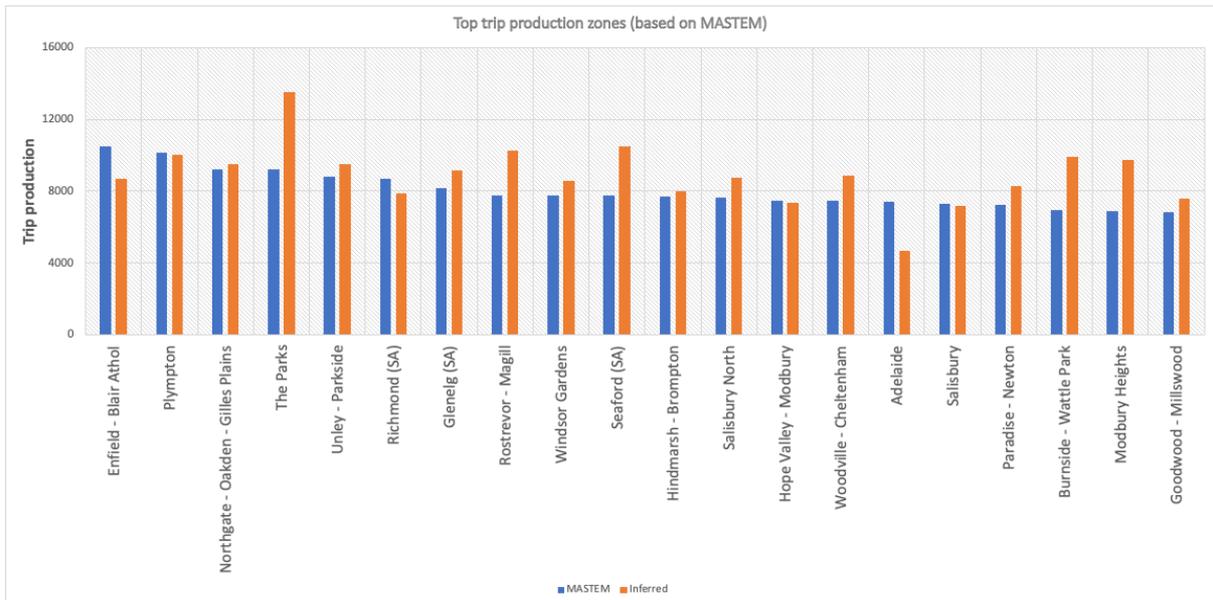

**Figure 4:** Comparison between daily zone-level trip production rates, as predicted by our inference algorithm (orange), and the current version of MASTEM (blue)

# 5. Sensitivity test and robustness analysis

In Section 5.1, we assess the stability of our inference results to the choice of weights $w_j$ and $w_k$ assigned to different data sources. In Section 5.2, we assess stability as a function of the starting values for the inference algorithm. Finally, in Section 5.3, we assess stability as a function of the input data sources.

## 5.1 Sensitivity to data quality weights

The weights $w_j$ and $w_k$ in our inference algorithm denote the quality of the data sources $j$ and $k$. While the analyst will usually have access to some high-level information about the relative quality of different data sources, translating these into numerical weights is a somewhat subjective process. For example, for our case study context of Adelaide, we are confident that SCATS sites offer the highest data quality, followed by the Census, the Bluetooth sensors, and the HERE Technologies' in-vehicle navigation systems. However, the absolute weights assigned to each of these sources is still somewhat subjective.

We examine how inference results change as a function of changes in these weights. Table 2 assesses similarity between the inferred ODM and the MASTEM ODM for different weight configurations. In general, the total number of trips is within a 10 per cent range across different optimisation runs, and the correlation coefficient and cosine similarity between the two matrices is fairly stable as well. The sensitivity tests indicate that as long as the relative ordering of weights is consistent across different optimisation runs, the inferred ODMs are relatively stable as well.

**Table 1:** Weightings and results for the sensitivity test in the AM peak period

| Weights | | | | Total no. of inferred trips | Comparison with MASTEM ODM | |
|---|---|---|---|---|---|---|
| SCATS | Census | Bluetooth | HERE | | Correlation Coefficient | Cosine Similarity |
| 100 | 70 | 40 | 20 | 552,634 | 0.86 | 0.87 |
| 100 | 70 | 50 | 20 | 561,877 | 0.84 | 0.85 |
| 100 | 70 | 60 | 20 | 570,896 | 0.82 | 0.84 |
| 100 | 70 | 40 | 30 | 552,714 | 0.86 | 0.87 |
| 100 | 70 | 50 | 30 | 561,980 | 0.84 | 0.85 |
| 100 | 70 | 60 | 30 | 570,918 | 0.82 | 0.84 |
| 100 | 70 | 50 | 40 | 561,971 | 0.84 | 0.85 |
| 100 | 70 | 60 | 40 | 570,909 | 0.82 | 0.84 |
| 100 | 75 | 50 | 30 | 556,375 | 0.85 | 0.86 |
| 100 | 80 | 40 | 20 | 543,246 | 0.87 | 0.89 |
| 100 | 80 | 50 | 20 | 551,265 | 0.85 | 0.87 |
| 100 | 80 | 60 | 20 | 559,124 | 0.84 | 0.85 |
| 100 | 80 | 40 | 30 | 543,353 | 0.87 | 0.89 |
| 100 | 80 | 50 | 30 | 551,267 | 0.85 | 0.87 |
| 100 | 80 | 60 | 30 | 559,229 | 0.84 | 0.85 |
| 100 | 80 | 50 | 40 | 551,427 | 0.85 | 0.87 |
| 100 | 80 | 60 | 40 | 559,572 | 0.84 | 0.85 |

## 5.2 Sensitivity to starting values

As mentioned in Section 3, the target function is nonconvex and likely to have multiple local minima. Consequently, estimation results could potentially be sensitive to starting values, such that the inferred ODM does not reach the global optimum, and the optimisation algorithm ends up at a local optimum. To examine the stability of inference results, we repeated the inference 20 times using random starting values, and compared the inferred ODM to the one reported in the previous Section, where the starting values for the ODM were chosen to be the ODM values observed within the Census (because the Census is our most reliable source of ODM information).

We find the inference results to be remarkably stable. Across our 20 runs, the highest total inferred number of trips is 556,544, and the lowest total number is 556,345, compared to 556,374 for the reported result. From a structural perspective, we calculated the cosine similarity and correlation coefficient between each pair of non-identical results, and in all cases both measures were higher than 0.99. In general, we find that despite the existence of non-convexity in the target function, the proposed inference framework can produce stable results, and is not very sensitive to starting values.

## 5.3 Sensitivity to input data sources

A major advantage to our inference framework is its ability to incorporate movement and count information from any number of different data sources with varying levels of data quality. We demonstrate the benefits of this flexibility by assessing the loss in inference quality if we were to omit information from the Census.

In the absence of Census data, we infer total number of trips during the AM peak period to be roughly 783,000, about 41 per cent higher than our inferences when we included the Census ODM, and 62 per cent higher than the MASTEM ODM. From a structural perspective, the cosine similarity between the inferred ODM without Census data and the MASTEM ODM drops to 0.37, and the correlation coefficient drops to 0.36. For trip productions, the cosine similarity between the two matrices is 0.74, and the correlation coefficient is 0.45. For trip attractions, the cosine similarity is 0.79, and the correlation coefficient 0.62. In general, these findings indicate that the inferred ODM is much less similar to the MASTEM ODM in the absence of Census data.

Note that the Census is the second most reliable source of information in our inference algorithm. Therefore, the loss in inference quality is high when this information is excluded from the inference framework. In general, the improvement in inference quality from including an additional source of information will be proportional to the quality of the new information source. Our analysis indicates that the inclusion of new sources of high-quality information can appreciably improve inference results.

# 6. Conclusions

This study developed a methodology for inferring OD flows within a planning area that does not require data collected actively from household travel surveys. Instead, the methodology uses passively collected mobility information from IoT and other existing sources. The proposed framework is flexible in that it can incorporate any kind of movement or count data to infer OD flows for any transport mode, while controlling for differences in data quality across different sources. The proposed framework is scalable in that it can readily be operationalised for large realistic networks and urban areas.

The framework is applied to infer OD flows for car travel within the metropolitan area of Adelaide, Australia, using data from loop detectors, roadside Bluetooth sensors, in-vehicle navigation systems and the Australian Census. We compare our OD inferences with those from the current version of the MASTEM – Adelaide's strategic transport demand model, calibrated using data from traditional household travel surveys. We find remarkable consistency between our inferences and those from MASTEM, despite differences in input data and methodologies. For example, for the AM peak period, we predict the total number of trips made within Adelaide to be equal to 556,000, while the corresponding prediction from MASTEM is equal to 484,000. The two predictions are within 20 per cent of each other. When we compare the spatial distribution of trips, in terms of origins and destinations, we find that our inferred OD matrix has an 86 per cent cosine similarity to the corresponding MASTEM matrix. In summary, our results show that the proposed framework can produce highly comparable ODM, trip production and trip attraction patterns to those inferred from traditional household travel survey-based transportation demand modelling methods.

In general, our real-world experiment demonstrates the effectiveness of the proposed ODM inference framework. By integrating information from multiple data sources (four in the experiment), we are able to control for missing variables, measurement errors and other sources of systematic bias that might apply to any one single information source. Consequently, our inferences are less likely to be distorted by data sources with relatively lower quality, and more likely to deliver reasonable and realistic results. By adopting a single-level framework for converting OD flows into equivalent link-level counts, we are able to apply the framework to large realistic networks and urban areas with several thousand OD pairs. For example, our inference algorithm takes roughly 10 minutes to converge on a personal computer for a region with roughly 100 OD zones, and 10,000 OD pairs.

A key limitation to the proposed framework is that the spatial level of our analysis is necessarily limited by the availability of appropriate information. For our case study region of Adelaide, the density of loop detectors and road-side sensors was not high enough to support direct application of our framework to infer OD flows at a TAZ level. Ongoing work is exploring ways in which outputs from our inference framework could potentially be interpolated to impute OD flows at higher levels of spatial resolution.